\documentclass[a4paper]{article}
\textheight22truecm \usepackage{t1enc}\usepackage[latin1]{inputenc}
\usepackage[english]{babel}\usepackage{graphicx}\usepackage{layout}
\begin{document}
\begin{center}
{\Large \bf Prediction of the human life expectancy }
\end{center}
\vspace{0.5cm}
\begin{center}
{\large \sl A. {\L}aszkiewicz,  Sz. Szymczak, S. Cebrat${\it ^1}$}\\

\noindent Department of Genomics, Institute of Genetics and Microbiology,
University of Wroc{\l}aw, ul. Przybyszewskiego 63/77, PL-54148 Wroc{\l}aw,
Poland, \\
\noindent cebrat@microb.uni.wroc.pl
http://smORFland.microb.uni.wroc.pl\\

\end{center}
{\it $^1$ To whom all correspondence should be sent.}
\vspace{0.5cm}
\hrule
\vspace{0.5cm}
\noindent{\bf Abstract}\\
\noindent
We have simulated demographic changes in the human population using the Penna
microscopic model, based on the simple Monte Carlo method. The results of
simulations have shown that during a few generations changes in the genetic pool
of a population are negligible, while improving the methods of compensation of
genetic defects or genetically determined proneness to many disorders
drastically affects the average life span of organisms. Age distribution and
mortality of the simulated populations correspond very well to real
demographic data available from different countries. Basing on the comparison of
structures of real human populations and the results of simulations it is
possible to predict changes in the age structure of populations in the future.
\\


\noindent{\bf Introduction}\\
\noindent
Human life expectancy has increased during the last century significantly. For
instance, chance for American people at birth to survive until 80 years
increased almost fourfold during that time (data available at
http://www.mortality.org) \cite{mortality}. No doubt, it is a great achievement of our
civilisation. On the other hand such deep changes in the age structure of the
human population have caused a lot of problems for many social institutions. For
example, insurance companies signing agreements with their clients for the
retirement rents are interested in correct predictions of such demographic
changes. In fact it is important to know not only the average life expectancy
for the future generations but also the age distributions of populations. The
higher mortality of newborns influences strongly the life expectancy at birth
while it can have only slight effect on the life expectancy at the age of $20$.
Since one century corresponds to only a few generations in the human population,
it is a too short period for a significant reconstruction of its genetic pool.
Thus, there should be some effects other than genetic influencing the ageing
processes and the life span of humans. Recently Stauffer has described a
phenomenological model for the prediction of the human mortality in rich
countries \cite{Stauffer}. Below we try to show that a simple microscopic Penna model \cite{Penna},
based on the Monte Carlo method can be useful for simulating the processes of
demographic changes which have occurred during the last century. Due to one
simple assumption that at least some genes in the genomes are
switched on chronologically, the Penna model predicts that the selection against
defective genes expressed before the minimum reproduction age is much stronger
than the selection against genes expressed after the minimum reproduction age.
Furthermore, there is a gradient of selection pressure for the late-expressed
genes - stronger selection for genes expressed just after the minimum
reproduction age and weaker selection for genes switched on during the older
ages. This generates a gradient of fractions of defective genes in the
genomes - a higher fraction of defective genes are expressed in the late ages. Such an uneven distribution of defective genes in the genomes is responsible for the specific age structure of the simulated populations.
We have assumed that it should be possible to simulate the changes in the age
distribution observed in the natural human populations just by changing the
tolerance for the number of expressed defective characters. This mimics the
changes in the human life style and the amelioration of medical care. Many
defects which used to be lethal in the past now do not kill people inevitably.
We have learnt how to produce a phenocopy - a person who looks and lives like a
healthy individual, as in the case of phenyl-ketonuria or haemophilia. It is
possible to avoid an onset of a fatal disorder, for example cancer, which
could be promoted by a genetically transferred recessive gene - a suppressor of
oncogenes. There are even more drastic situations - for example, due to the technical achievements and the progress in endocrinology people without functional kidneys can live and be active for a very long time.
The influence of the number of tolerated defective characters on the age distribution was analysed previously. P.M. de Oliveira et al., \cite{Paulo} simulated evolution of populations using Penna model where individuals were represented by bitstrings 32 bits long. Such short "genomes" produced too big shifts in the maximum age of populations. Niewczas et al.,  \cite{Ewa} simulated the effect of changes in the tolerance for the genetic defects allowing the evolution of populations for about 10 - 20 generations. Below we show that it is possible to simulate the changes in the age distribution of populations without any change in their genetic pool, giving no time for the population evolution.
\\

\noindent{\bf Model}\\
There are many modifications of the standard Penna model, see \cite{War} for review. In our
version, in the sexually reproducing population each individual is represented
by a diploid genome. Each genome is composed of two bitstrings (haplotypes), each 640 bits long. A mutated gene (set for $1$) can be complemented by its wild allele (set for $0$ - correctly executing its function) which means that all mutations are recessive. There are no reversions in the model - the mutation replaces the wild allele by a defective one but a defective allele is not
replaced by a wild one and stays defective after the mutation process. Genes are
switched on chronologically, which means that during each individual's life
span, the genes at the same loci are switched on at the same age. Furthermore,
as in the standard model, the number of genes switched on grows linearly with the age of the organism but it is easy to modify the model in such a way that the number of genes switched on during the first stages of embryonic development is much larger \cite{House}, which produces higher mortality during pregnancy, as estimated for the human population \cite{Hassold},\cite{Copp}. The
assumption of the model that at least some genes are switched on
chronologically, produces a gradient of frequency of defective genes which results in higher incidence of defective phenotypes in the older
ages (higher frequency of homozygotes with both defective alleles at the locus).
There is another important assumption in the model - individuals have a declared
tolerance to the number of expressed defective phenotypic characters which
means that not every expressed defect kills an individual. In the model, usually
this tolerance is set for 3 which means that the third expressed defect
eliminates the individual from the population.
At each time step two alleles at the consecutive locus are
switched on. If in the switched on loci, at three positions, both alleles are
defective - the individual dies (threshold $T=3$). When the $200th$ bit is switched on, the individual reaches its reproduction age. A female at the reproduction age produces a gamete with a probability $0.25$. The gamete is a
product of one cross-over between two haplotypes of the parental genome, in
corresponding sites. After cross-over, one locus in a gamete is chosen for
mutation and if it is $0$ - it is replaced by $1$, if it is $1$ - it stays $1$. A newborn is formed by joining this gamete with another one produced in the same
way by a randomly chosen male individual, also at the reproduction age. To avoid
unlimited growth of the population, the Verhulst factor  $V$ is introduced: ${V=1-N_{t}/N_{max}}$ where ${N_{max}}$ - the maximum
population size - is often called the capacity of the environment, and $N_{t}$  is the current population size. For each zygote a random number between $0$ and $1$ is generated and if it is greater than $V$, the zygote
dies. In our simulations the Verhulst factor operates only at conception, meaning zygote formation, which implies that there are no random deaths in the
population \cite{SaMartin}. The maximum capacity of the environment is set for $50 000$.
To show how the higher tolerance for the genetic defects affects the age distribution of the populations we have used two variations of the model.\\

1. We simulated the evolution of populations under the threshold $T=3$ and after $80 000$ MC steps we produced one million zygotes. Then we anticipated the life span of each zygote under a different threshold $T$. In particular, we anticipated the life spans under $T$ which were not integers. In such cases, the probability of surviving under a given $T$ is equal to
a fractional part of $T$, which means that  individuals under threshold $T=3.3$ have a probability $0.7$ to die when three bad mutations are reached.\\

2. After $80 000$ MC steps we gradually increased the
threshold $T$ from $3$ to $5$ during $800$ MC steps (by $0.1$ each 40 MC steps).
\\

\noindent{\bf Results and Discussion }\\
The results of simulations of populations with parameters described above are
shown in Fig. $1$. The different plots in the Figure represent the anticipated age distribution of individuals developed from zygotes produced by the population which evolved under threshold $T=3$ if they lived under different threshold $T$. Note, that
numbers on the x-axis correspond to the numbers of the consecutive loci in the bitstring and - simultaneously - to the age of individuals. As it was mentioned
above, this modification of the model does not predict the higher prenatal or
newborns' mortality. It is possible to model this period of the human life span
by assuming that during the first stages of development a much larger fraction
of genes is switched on than during the period after reaching the minimum
reproduction age \cite{House}. This simplification does not change significantly the other predictions of the model. \\
\noindent
In this paper we have concentrated on the age
distribution of the population after it reaches the minimum reproduction age ($200th$ bit). To rescale the x-axis we assumed that the minimum reproduction age corresponds to 20 years in the real human population and the maximum age in the simulated population under threshold $T=3$ corresponds to the maximum age of the real human
population at the end of the $19th$ century (American population in the
period $1870-1879$).
After such a rescaling we got the plot shown in Fig. $2$. In this Figure a
series of plots illustrating the age distributions of the American populations in
different periods from the end of $19th$ century to the end of the $20th$ century is also shown. Note, that the computer
simulated population with the parameters described above ($T=3$) fits to the human real population at $1870 - 1879$. Nevertheless, the shape of the age distribution of
the real human populations tens of years later underwent "rectangularization",
which means that the mortality of the younger fraction of the population decreased and a much
larger fraction of the whole population survived until the older ages. The same
effect has been obtained in the computer simulated populations just by
assuming that individuals are less prone for deleterious effects of genetic
defects (increasing threshold $T$, see also Fig. 1).
Data representing  the logarithm of mortality plotted against the age are
shown in Fig. $3$. This linear relation between the logarithm of mortality and
the age is known as Gompertz law. As shown in Fig. $3$, the differences between
populations from the beginning and the end of the $20th$ century mainly
correspond to the different slopes of the lines illustrating changes in the
mortality which has been shown and discussed many times, see \cite{Stauffer}, \cite{humanmortality}. \\
\noindent
The same effect has been observed in the second variation of the model, which assumes that populations is gradually less and less prone to the deleterious effects of the defective genes in
their genomes. In our simulations one generation corresponds to about $250$ bits
($200$ bits correspond to the minimal reproduction age). To show, that the
reconstruction of the genetic pool of the population under such condition is
negligible, we have performed the simulation under the gradually increasing
threshold $T$ from $3$ to $5$ during the $800$ MC steps.
Nevertheless, if we assume that the number of mutations in the real population
is of the order of $1$ per genome per generation \cite{Mukai}, \cite{Drake}, the effect of
mutation accumulation should be negligible. In fact, the differences in age distribution
and mortality between populations simulated under such conditions and previously
estimated are not significant (data not shown) what confirms previous results \cite{Ewa}. \\
\noindent
Results of our simulations indicate that the rate of increasing the threshold $T$
during the last century was not constant but the range of changes was
between $0.1 - 0.2$ $T$-unit per $10$ years. If we assume that this rate will not
change in the near future, it is possible to predict the further changes in the
human age distribution. Our simulations predict further rectangularization of
the curve describing the age distribution and a steeper slope in the Gompertz plot
with a rather small shift of the maximum life span (see the predictions for
$T=6.3$ in Figs. 3 and 4). Observations described by Yashin et al \cite{Yashin} and predictions published at web sites \cite{mortality} suggest a rather parallel shift of the age
distribution curve toward the higher ages with a relatively high shift in the
maximum life span and no further changes in the slope of Gompertz curve.
Our predictions give higher life expectancy at lower ages but lower life
expectancy estimated for people who have already reached a higher age (Fig. 4).\\

\noindent
In conclusion - the microscopic Penna model of the evolution of the age
structured populations based on very simple assumptions concerning the genetic
structure and function of individuals gives reasonable results which could be
considered by demographers. Furthermore, a better understanding of the model can give much better insight into the biological significance of its parameters.\\
\noindent

\noindent{\bf Acknowledgemets}\\ The authors thank D. Stauffer for
comments and discussions. The work was supported by the grant number
1016/S/IMi/03.\\

%\noindent{\bf Figures captions}

%\noindent
\begin{figure}[hbt]
\begin{center}
\includegraphics[angle=-90,scale=0.6]{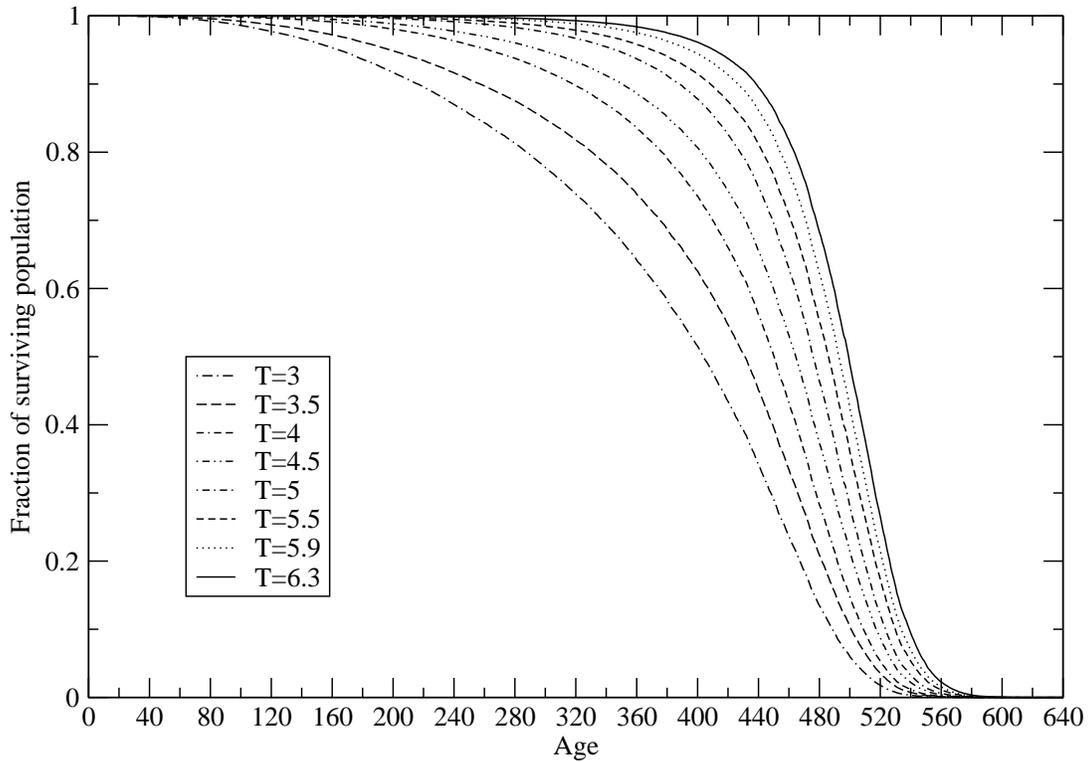}
\end{center}
\caption{
Age distributions of populations which evolved at threshold $T=3$ and were checked for survival probability under different thresholds $T$ (see text for more explanation). Age of individuals (x-axis) scaled in the number of consecutive loci in the bitstring corresponding to the age in MC steps.
}
\end{figure}
%\noindent
\begin{figure}[hbt]
\begin{center}
\includegraphics[angle=-90,scale=0.6]{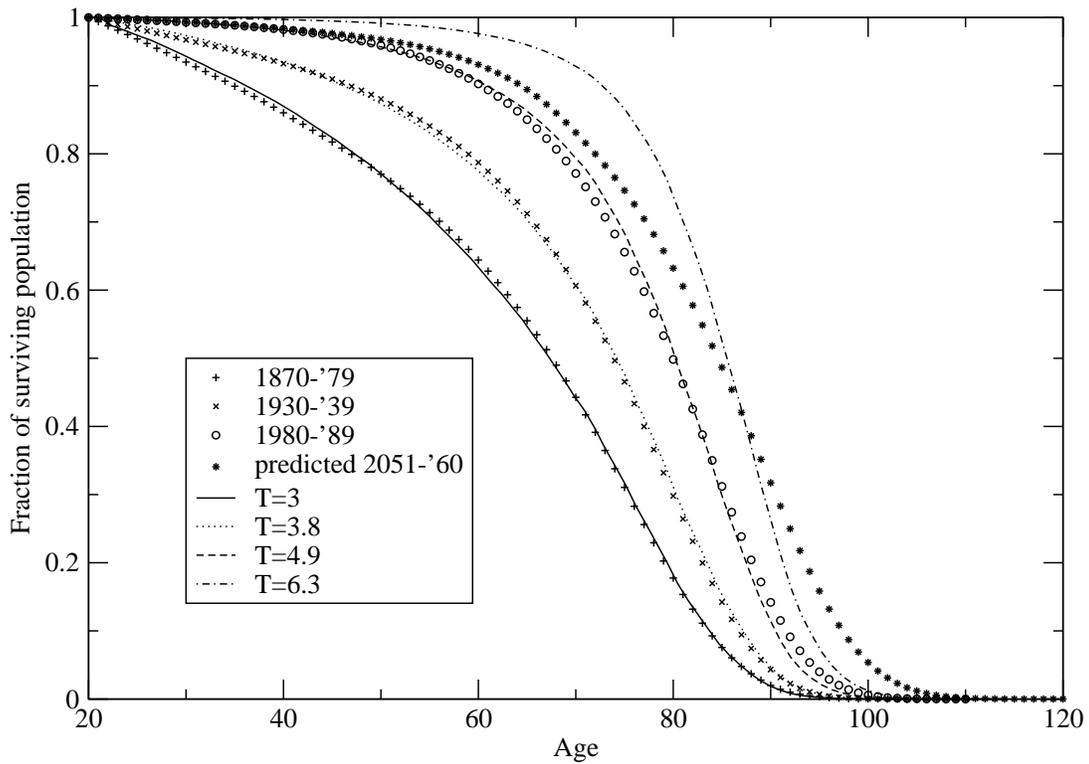}
\end{center}
\caption{
Fitting the age distributions of American populations from different periods to the age distributions of the simulated populations. The x-axis values of the last ones were rescaled as described in the text.
}
\end{figure}
%\noindent
\begin{figure}[hbt]
\begin{center}
\includegraphics[angle=-90,scale=0.6]{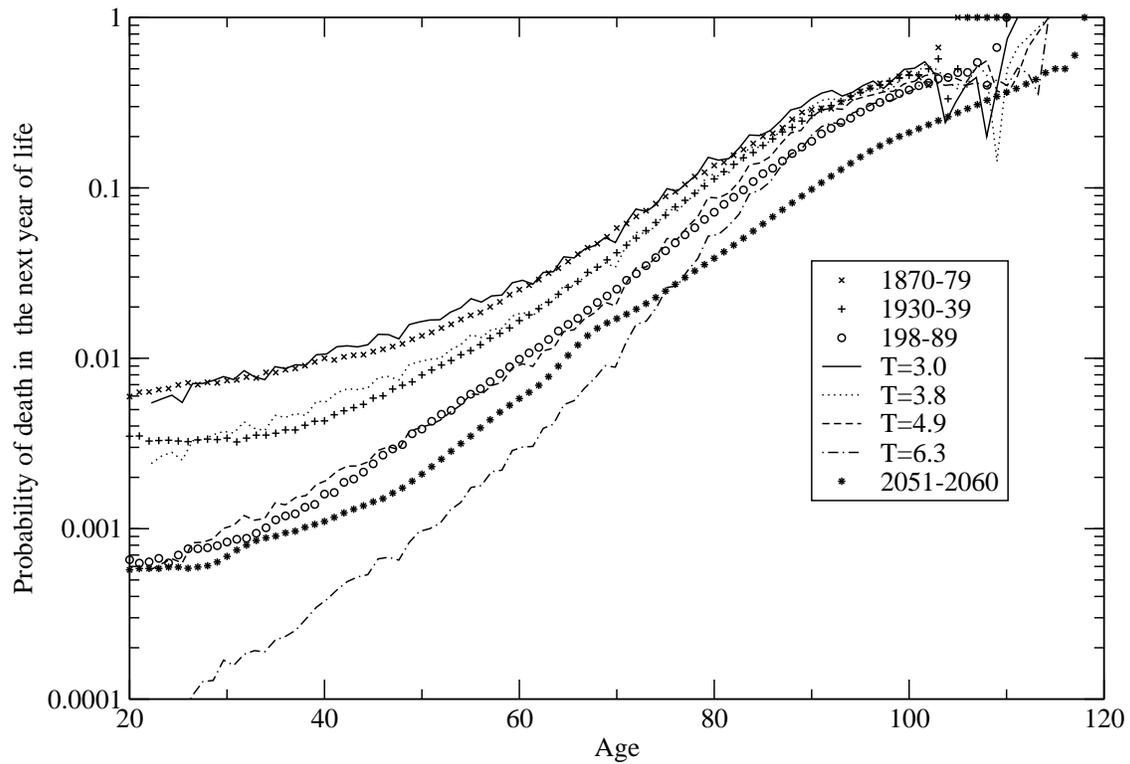}
\end{center}
\caption{
Gompertz curves for the natural American populations and for the simulated populations. Note the different slopes for predictions for years 2051-60 and for the simulated population under $T=6.3$.
}
\end{figure}
%\noindent
\begin{figure}[hbt]
\begin{center}
\includegraphics[angle=-90,scale=0.6]{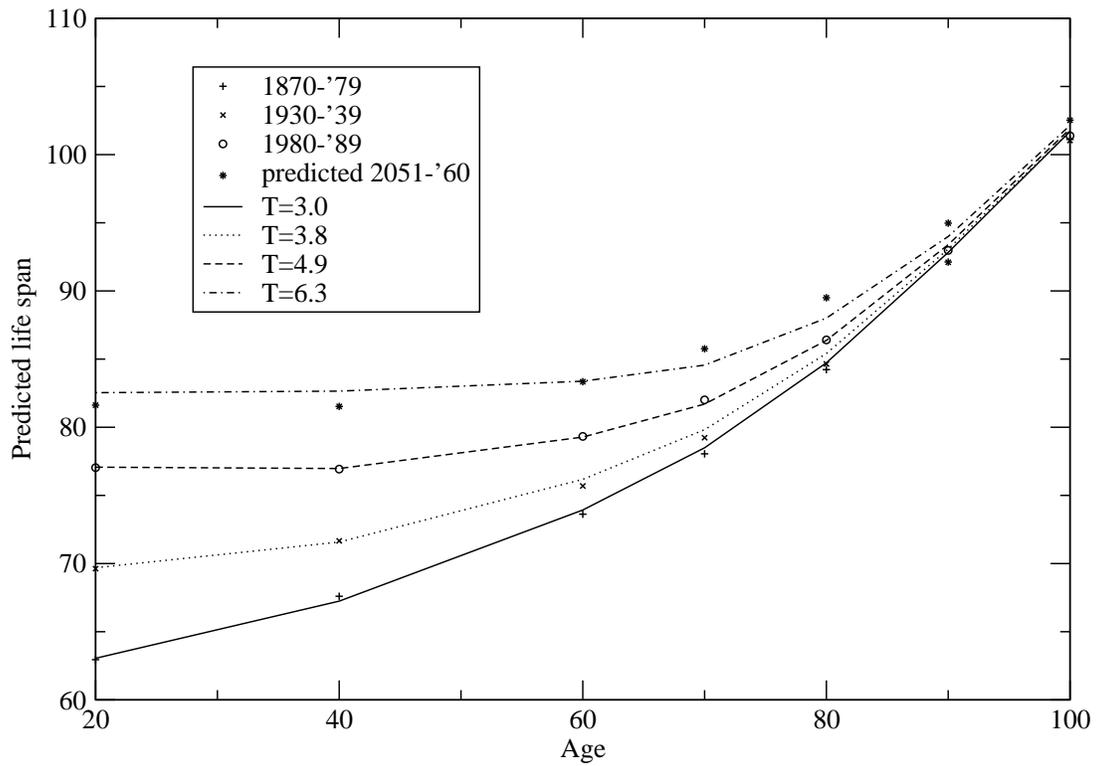}
\end{center}
\caption{
 Predicted life span at different ages for natural American populations and for simulated populations. "Predicted" for the ancient American populations means what was the life span of the part of population which reached the given age (in fact just calculated, not predicted).
}
\end{figure}

\end{document}